\documentclass[10pt]{IEEEtran}

\usepackage{shortcuts_tmr}

\def\cH{{\cal H}}
\def\A{{\bm A}}

\def\I{{\bm I}}

\def\Cy{{\bm C}_y}
\def\Cym{\widehat{\bm C}_y^m}

\def\wl{{\bm w}^{\Paren{\ell}}}
\def\yl{{\bm y}^{\Paren{\ell}}}

\def\u{{\bm u}}
\def\uhat{\widehat{\u}}
\def\v{{\bm v}}
\def\vhat{\widehat{\v}}

\def\x{{\bm x}}

\def\beps{{\bm \epsilon}}

\def\G{{\cal G}}
\def\Verts{{\cal V}}
\def\Edges{{\cal E}}

\def\CONE{{\cal C}}
\def\BOUNDARY{\partial\CONE}

\title{Blind Estimation of Eigenvector Centrality from Graph Signals: Beyond Low-pass Filtering}
\author{
  T.~Mitchell~Roddenberry and Santiago~Segarra
  \thanks{
    T.M. Roddenberry and S. Segarra are with the Dept. of ECE, Rice University.
    TMR partially received funding from the Ken Kennedy 2019/20 AMD Graduate Fellowship.
    Emails: mitch@rice.edu, segarra@rice.edu.
  }
}

\begin{document}

\maketitle

\begin{abstract}
  This paper characterizes the difficulty of estimating a network's eigenvector centrality only from data on the nodes, \ie with no information about the topology of the network.
We model this nodal data as graph signals generated by passing white noise through generic (not necessarily low-pass) graph filters.
Leveraging the spectral properties of graph filters, we estimate the eigenvectors of the adjacency matrix of the underlying network.
To this end, a simple selection algorithm is proposed, which chooses the correct eigenvector of the signal covariance matrix with minimal assumptions on the underlying graph filter.
We then present a theoretical characterization of the asymptotic and non-asymptotic performance of this algorithm, thus providing a sample complexity bound for the centrality estimation and revealing key elements driving this complexity.
Finally, we illustrate the developed insights through a set of numerical experiments on different random graph models.

\end{abstract}

\section{Introduction}\label{sec:intro}

The representation of data as graphs, or networks, has become an increasingly prominent approach in science and engineering~\cite{newman2010,jackson2010}, allowing one to uncover community structure~\cite{newman2006modularity}, common connection patterns~\cite{milo2002networkmotifs}, and node importance~\cite{pagerank1999}.
In many settings, there is an assumed network structure lying underneath a set of interacting agents, but the precise connections in this structure are unobserved.
However, we would still like to use graph-based analysis tools, such as centrality measures~\cite{beauchamp1965improved,freeman1977set,segarra2016stability,bonacich1972factoring}, to draw conclusions about the role of the agents in the unobserved interconnected structure.

As a motivating example, consider a social network where the set of connections is not known precisely, but one has measurements of opinion dynamics among all of the individuals.
One then seeks to infer who the most influential individual in the network is.
In the network topology inference framework, one would use this collection of opinions to infer a graph structure, and then compute the eigenvector centrality of the constructed network.
This approach requires the costly -- both in the data and computational sense -- construction of an intermediate network, even though the ultimate interest is just in the resulting centrality.
This leads to the guiding question of this work:
\emph{How can we estimate the eigenvector centrality of a graph with hidden edges directly from data supported on the nodes?}
Working in the framework of graph signal processing, we model this data as a set of graph signals, obtained via the output of a graph filter applied to white noise.
We then explore the difficulty of this problem, characterized by the distribution of centrality over the nodes as well as the spectral properties of the graph and graph filter.

\noindent\textbf{Related work.}
One can frame the general problem of inferring the complete set of edges of an underlying network from data on the nodes under the concept of \emph{network topology inference}.
This is commonly studied through a statistical lens, where each node represents a random variable, and edges encode the dependence structure between these random variables~\cite{friedman2007lasso,lake2010discovering,meinshausen2006high}.
Alternative approaches include those based on partial correlations~\cite{friedman2007lasso}, structural equation models~\cite{bazerque_2013_sparse,baingana_2014_proximal}, and Granger causality~\cite{sporns_2012_discovering}.
Graph signal processing provides a different view on network topology inference, where the nodal data is assumed to be the output of a latent process on the hidden graph~\cite{dong2016laplacian,kalofolias2016smooth,segarra2017topo,mateos2019connecting}.
These approaches solve the topology inference problem by making different assumptions on the process that generates the graph signals, \eg kernel models~\cite{shen2017kernel}, signal smoothing~\cite{dong2016laplacian}, or consensus dynamics~\cite{segarra2017consensus, zhu2019network}.

Motivated by the high sampling and computational requirements of network topology inference, the framework of \emph{blind network inference} was proposed.
More precisely, recent works have considered the estimation of network characteristics -- such as community structure and centralities -- directly from nodal data, \ie circumventing the intermediate step of constructing the network.
In this context, \cite{schaub2018dynamical} considers the observation of a simple finite-length diffusion process with white noise input on a planted partition graph, and characterizes the relationship between the diffusion time and the difficulty of identifying the latent communities from the observed graph signals.
Alternatively, \cite{wai2018lowrank} models the observed signals as being low-rank, while \cite{hoffmann2018unobserved} models them as a function of a latent time-series.
Expanding on this, \cite{roddenberry2020community} considers the blind community detection problem where there is no fixed graph, but rather a family of graphs with shared latent partitions.
Most related to this work, \cite{roddenberry2020ranking,he2020centrality} implement this blind inference approach for the estimation of eigenvector centralities from graph signals.
Both of these works assume that the graph signals are smooth, and characterize the error between the estimated and true centralities. We depart from this assumption in the current paper.

\noindent\textbf{Contributions.}
The contributions of this paper are three-fold:
\begin{enumerate*}
\item We provide a simple algorithm to select an estimate of the eigenvector centrality from a set of graph signals,
\item We derive sampling requirements for this algorithm, and
\item We illustrate our theoretical findings through experiments on different random graph models.
\end{enumerate*}

\section{Preliminaries}\label{sec:prelim}

\subsection{Graphs and eigenvector centrality}\label{sec:prelim:graphs}

An undirected \emph{graph} $\G$ consists of a set $\Verts$ of $n\deq\abs{\Verts}$ nodes, and a set $\Edges\subseteq\Verts\times\Verts$ of edges, corresponding to unordered pairs of nodes.
Networks can be represented by an \emph{adjacency matrix} $\A$ defined by first setting an arbitrary indexing of the nodes with the integers $1,\ldots,n$, and then assigning $A_{ij}=1$ if $\Paren{i,j}\in\Edges$ and $A_{ij}=0$ otherwise.

The \emph{eigenvector centrality} of a node $i$ in a graph $\G$ is given by the $i^{\rm th}$ entry of the leading eigenvector of $\A$, which we denote $\u$ for convenience.
The Perron-Frobenius Theorem~\cite[Theorem 8.3.1]{horn2012matrix} guarantees that every element of $\u$ has the same sign.
Due to this property, the set of all possible eigenvector centralities (ignoring the stipulation of having unit norm) can be characterized as a symmetric convex cone $\CONE = \CONE^+ \cup \CONE^- \in \R^n$ where
\begin{equation*}\label{eq:centrality-cone}
    \CONE^+ = \Brack{\x\in\R^n \colon x_i\geq 0}, \,\,\, \CONE^- = \Brack{\x\in\R^n \colon x_i\leq 0}. 
\end{equation*}
The boundary of this set, then, is
\begin{equation*}\label{eq:centrality-cone-boundary}
  \BOUNDARY = \Brack{\x\in\CONE \colon \prod_{i=1}^n x_i=0}.
\end{equation*}
That is, $\BOUNDARY$ consists of vectors with same-signed entries and at least one entry that takes value $0$, while $\CONE$ consists of vectors with same-signed entries.
In fact, for connected, undirected graphs, the leading eigenvector of $\A$ lies in the interior of $\CONE$, since no node can have centrality exactly equal to $0$.
Of interest in this work is the projection of a vector $\v$ onto $\CONE$, denoted by $\Proj{\CONE}{\v}$, which essentially takes the positive-signed or negative-signed part of $\v$, depending on which is closer to $\v$ in the $\ell_2\text{-norm}$.

\subsection{Graph signals and graph filters}\label{sec:prelim:gsp}

\emph{Graph signals}, analogously to discrete time signals, are functions mapping the nodes to the reals, \ie $x\colon\Verts\to\Real$.
For an indexing of $\Verts$ with $\Sbrack{n}$, a graph signal $\x$ is represented as a vector in $\Real^n$, where $x_i=x\Paren{i}$.
A graph filter $\cH$ is a linear map between graph signals representable as a polynomial of the adjacency matrix\footnote{Alternatively, graph filters can be defined in terms of other graph matrices, such as the Laplacian matrix. We focus on adjacency-based filters.}
\begin{equation}\label{eq:graph-filter}
  \cH\Paren{\A} = \sum_{k=0}^T\gamma_k\A^k \deq \sum_{k=0}^T\cH\Paren{\lambda_k}\v_k\v_k^\top,
\end{equation}
where $\gamma_k$ are real-valued coefficients, $\cH\Paren{\lambda}$ is the extension of the polynomial $\cH$ to scalar-valued arguments, and $(\lambda_k,\v_k)$ denote the eigenpairs of $\A$.

\subsection{Blind network inference}\label{sec:prelim:blindinference}

Recent works have considered the problem of \emph{blind network inference}, where one aims to infer coarse network descriptions, such as the community structure~\cite{roddenberry2020community,Schaub2019,hoffmann2018unobserved} or eigenvector centrality~\cite{roddenberry2020ranking,he2020centrality} solely from graph signals, not knowing the underlying graph structure.
These works mostly rely on an assumed model for the observed graph signals: in essence, that the graph signals are the output of a low-pass filter applied to white noise.
That is, each observed graph signal $\yl,\ 1\leq\ell\leq m$ can be written as
\begin{equation}\label{eq:system-model}
  \yl \deq \cH\Paren{\A}\wl,
\end{equation}
where $\cH$ is a polynomial of the graph adjacency matrix $\A$, and the set $\Brack{\wl}_{\ell=1}^m$ consists of \iid\ samples from a zero-mean distribution obeying $\Cov\Paren{\wl} \deq \E\Sbrack{\wl\Paren{\wl}^\top}=\I$.
Unlike existing works, we do not assume the filter $\cH$ in \eqref{eq:system-model} to be low-pass. 

By considering \eqref{eq:graph-filter}, the covariance matrix of signals following~\eqref{eq:system-model} shares a set of eigenvectors with the adjacency matrix.
Specifically,
\begin{equation}\label{eq:shared-eigenvectors}
  \begin{split}
    \Cy &\deq \Cov\Paren{\yl} = \cH\Paren{\A}\Cov\Paren{\wl}\cH\Paren{\A} \\
    &\,= \Sbrack{\cH\Paren{\A}}^2 = \sum_{k=0}^T\Sbrack{\cH\Paren{\lambda_k}}^2\v_k\v_k^\top.
  \end{split}
\end{equation}
Consequently, one can study the eigenvectors of the covariance matrix $\Cy$ to gain insights into the spectral structure of the adjacency matrix $\A$.

\section{Problem Statement and Algorithm}\label{sec:algs}

Consider a set of $m$ graph signals obtained as the output of an \emph{unknown} graph filter, as in~\eqref{eq:system-model}.
Then, based on~\eqref{eq:shared-eigenvectors}, one can analyze the spectral structure of a graph strictly from the observation of such signals, without knowledge of the graph itself.
More precisely, we aim to extract the best estimate of the eigenvector centrality $\u$ from the eigenvectors of the sample covariance.
This leads to the \emph{eigenvector centrality selection} problem, stated next.
\begin{problem}\label{prob:selection}
	Given the observation of $m$ graph signals following the model in~\eqref{eq:system-model}, estimate the eigenvector centrality $\u$.
\end{problem}

\noindent Based on the shared set of eigenvectors between $\Cy$ and $\A$, it makes intuitive sense to estimate the eigenvector centrality by selecting an eigenvector from the sample covariance matrix.
However, due to noise induced by finite sampling and the fact that $\cH$ is unknown, it is not immediately clear which eigenvector should be chosen.

Given certain assumptions on the graph filter $\cH$, one could select the leading eigenvector of the empirical covariance matrix 
\begin{equation}\label{E:sample_covariance}
\Cym = \frac{1}{m} \sum_{\ell=1}^m (\yl - \bar{\bm y}) (\yl - \bar{\bm y})^\top
\end{equation}
as an estimate of $\u$, as done in~\cite{roddenberry2020ranking,he2020centrality}.
In this work, we make no assumptions on the structure of the graph or the graph filter, hence,  the position of the optimal estimate $\uhat$ in the spectrum of $\Cym$ is unknown \apriori.
To this end, we propose~\cref{alg:selection} for the eigenvector selection task.
This algorithm leverages the property that the eigenvector centrality must lie in $\CONE$ by choosing the eigenvector of $\Cym$ that is either in or closest to $\CONE$.
\begin{algorithm}[tb]
  \caption{Eigenvector selection algorithm}\label{alg:selection}
  \begin{algorithmic}[1]
    \STATE {\bf INPUT:} Set of $m$ graph signals $\{\yl\}_{\ell=1}^m$ 
    \STATE Compute the covariance matrix $\Cym$ as in \eqref{E:sample_covariance}
    \STATE Compute the eigenvectors of $\Cym$, yielding $\Brack{\vhat_i}_{i=1}^n$
    \FOR{$i\in\Sbrack{n}$}
    \STATE $s_i=\cos\theta\Paren{\vhat_i,\Proj{\CONE}{\vhat_i}}$
    \ENDFOR
    \STATE $j=\argmax_i s_i$
    \STATE $\uhat=\vhat_j$
    \STATE {\bf OUTPUT:} Estimated eigenvector centrality $\uhat$   
  \end{algorithmic}
\end{algorithm}

\section{Theoretical Results}\label{sec:theory}

Before characterizing the behavior of~\cref{alg:selection}, we establish the following results on the sample covariance matrix for signals following~\eqref{eq:system-model}.
 \vspace{-2mm}
\begin{prop}\label{prop:convergence}
  If, for some $r>0$, $\norm{\yl}_2\leq r$ holds for a collection of signals $\Brack{\yl}_{\ell=1}^m$ observed according to~\eqref{eq:system-model}, their sample covariance matrix $\Cym$ satisfies the following with probability at least $1-\eta$:
  \begin{equation*}\label{eq:convergence}
    \norm{\Cym - \Cy}_2\leq C_0\sqrt{\log\Paren{\frac{1}{\eta}}\frac{r}{m}},
  \end{equation*}
  where $C_0\in\Theta\Paren{\norm{\Cy}_2}$.
\end{prop}
\Cref{prop:convergence} is stated in~\cite[Corollary 5.52]{vershynin2010introduction}, and characterizes the rate of convergence of the sample covariance matrix to the population covariance matrix.
Although this relies on the assumption that the norm $\norm{\yl}$ is bounded, these results can easily be generalized to the case where $\yl$ has a subgaussian distribution, following~\cite[Corollary 5.50]{vershynin2010introduction}.
Next, we describe the alignment of the eigenspaces of $\Cym$ and $\Cy$.
\begin{prop}\label{prop:alignment}
  Under the same conditions as~\cref{prop:convergence}, for two corresponding eigenvectors $\v_j,\vhat_j$ of $\Cy, \Cym$, respectively, the following holds with probability at least $1-\eta$:
  \begin{equation*}\label{eq:alignment}
    \sin\theta\Paren{\v_j,\vhat_j} \leq 2\frac{C_0}{\delta}\sqrt{\log\Paren{\frac{1}{\eta}}\frac{r}{m}},
  \end{equation*}
  where $\delta=\min\Brack{\lambda_j-\lambda_{j-1},\lambda_{j+1}-\lambda_j}$ is the population eigengap for $\v_j$.
\end{prop}
\Cref{prop:alignment} follows from~\cref{prop:convergence} and~\cite[Theorem 2]{yu2014daviskahanstat}.
With these results gathered, we now proceed to the statement and proof of our main result.

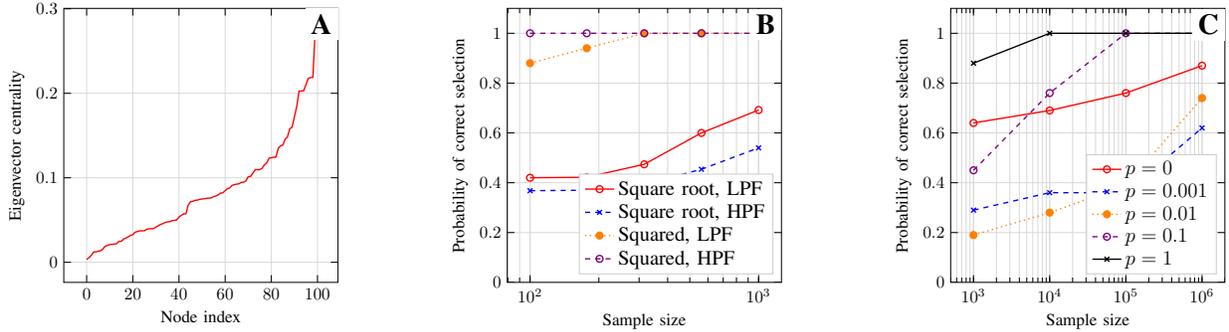
\begin{figure*}[t]
  \centering
  \resizebox{0.9\textwidth}{!}{\begin{tikzpicture}

  \begin{groupplot}[
    group style={
      group name=myplots,
      group size=3 by 1,
      horizontal sep=3.5cm},
    legend style={font=\large,draw=white!80.0!black},
    legend cell align={left},
    tick align=inside,
    tick pos=both,
    x grid style={gray!30},
    xmajorgrids,
    xminorgrids,
    xtick style={color=black},
    y grid style={gray!30},
    ymajorgrids,
    yminorgrids,
    ytick style={color=black},
    yticklabel style={
        /pgf/number format/fixed,
        /pgf/number format/precision=2
    },
    scaled y ticks=false,
    width=0.4\textwidth,
    height=0.4\textwidth
    ]

    \nextgroupplot[
      legend pos={north west},
      xlabel={Node index},
      ylabel={Eigenvector centrality},
      no markers,
    ]
    \addplot table[x=idx, y=u] {outdata/er-demo/centrality.csv};

    \nextgroupplot[
      legend pos={south east},
      xlabel={Sample size},
      ylabel={Probability of correct selection},
      ymin=0,
      xmode=log,
    ]
    \addplot table[x=m, y=correctness] {outdata/er-demo-sqrt-lpf/selection-correctness.csv};
    \addlegendentry{Square root, LPF}
    \addplot table[x=m, y=correctness] {outdata/er-demo-sqrt-hpf/selection-correctness.csv};
    \addlegendentry{Square root, HPF}
    \addplot table[x=m, y=correctness] {outdata/er-demo-sqrd-lpf/selection-correctness.csv};
    \addlegendentry{Squared, LPF}
    \addplot table[x=m, y=correctness] {outdata/er-demo-sqrd-hpf/selection-correctness.csv};
    \addlegendentry{Squared, HPF}

    \nextgroupplot[
      legend pos={south east},
      xmode=log,
      xlabel={Sample size},
      ylabel={Probability of correct selection},
      ymin=0,
    ]
    \addplot table[x=m, y=mean_correctness] {outdata/ws-p0000-hard/selection-correctness.csv};
    \addlegendentry{$p=0$}
    
    \addplot table[x=m, y=mean_correctness] {outdata/ws-p0001-hard/selection-correctness.csv};
    \addlegendentry{$p=0.001$}
    
    \addplot table[x=m, y=mean_correctness] {outdata/ws-p0010-hard/selection-correctness.csv};
    \addlegendentry{$p=0.01$}
    
    \addplot table[x=m, y=mean_correctness] {outdata/ws-p0100-hard/selection-correctness.csv};
    \addlegendentry{$p=0.1$}
    
    \addplot table[x=m, y=mean_correctness] {outdata/ws-p1000-hard/selection-correctness.csv};
    \addlegendentry{$p=1$}

  \end{groupplot}
  
  \node[below left,fill=white] at (myplots c1r1.north east) {\LARGE\textbf{A}};
  \node[below left,fill=white] at (myplots c2r1.north east) {\LARGE\textbf{B}};
  \node[below left,fill=white] at (myplots c3r1.north east) {\LARGE\textbf{C}};
\end{tikzpicture}

}
  \vspace{-3mm}
  \caption{
    Ranking algorithm for \ER (A,B) and \WS (C) random graphs, for~\cref{sec:experiments:erdos-renyi,sec:experiments:ws-selection}.
    (A) Eigenvector centrality of drawn \ER graph with $n=100,p={(\log{n})}/{n}$.
    (B) Rate of optimal eigenvector selection on \ER graph for $m\in\Brack{100,\ldots,1000}$.
    (C) Empirical probabilities of selecting the optimal estimate of the eigenvector centrality for 100 \WS random graphs with $n=500, k=4$.
  }
\vspace{-3mm}
  \label{fig:experiments:erdos-renyi}
\end{figure*}

\begin{figure*}[t]
  \centering
  \includegraphics[width=\linewidth]{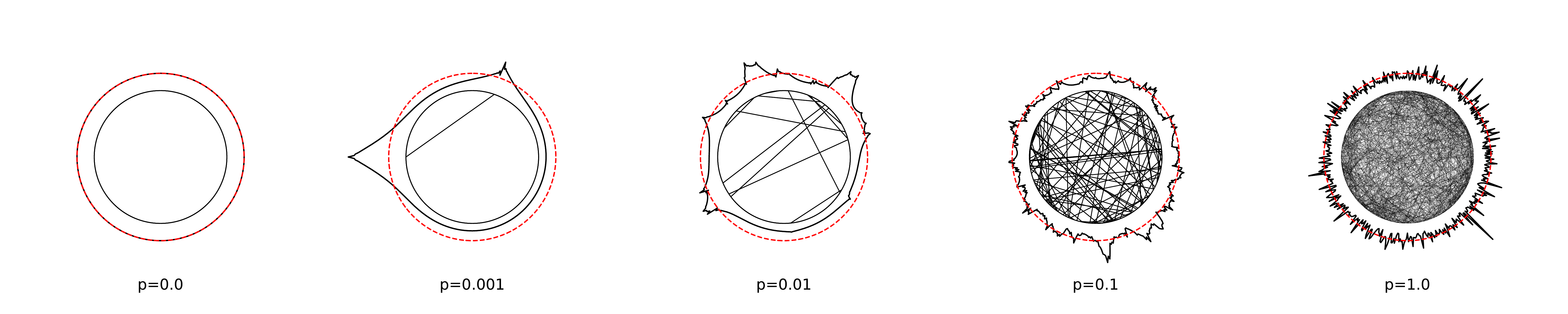}
  \vspace{-8mm}
  \caption{
    Eigenvector centralities of Watts-Strogatz random graphs with $n=500, k=4,$ and $p$ ranging from $0.0$ to $1.0$. Nodes are arranged on a circular grid and the centrality values are plotted around this circle for varying probabilities of rewiring $p$.
    For reference, the constant centrality of ${1}/{\sqrt{n}}$ is plotted with a red dashed line for each graph.
  }
\vspace{-4mm}
  \label{fig:experiments:ws-localization}
\end{figure*}

\vspace{-1mm}
\subsection{Eigenvector centrality selection guarantees}\label{sec:theory:selection}

Based on~\cref{prop:alignment}, it is clear that as $m\to\infty$, the eigenvectors of $\Cym$ align perfectly with those of $\A$, so~\cref{prob:selection} will be solved perfectly by simply selecting the only eigenvector that lies in $\CONE$, corresponding to the eigenvector centrality.
However, due to noise in $\Cym$, we aim to characterize how much data is needed to ensure correct selection of $\uhat$.
\begin{theorem}{\bf (Selection in the finite-sample regime)}\label{thm:selection}
  Let $j\in\Sbrack{n}$ be such that the $j^{\rm th}$ eigenvector of $\Cy$ is equal to $\u$.
  Define $\delta=\min\Brack{\lambda_j-\lambda_{j-1},\lambda_{j+1}-\lambda_j}$, as in~\cref{prop:alignment}.
  Then, if
  \begin{equation}\label{eq:finite-selection-requirement}
    m\in\Omega\Paren{\max_{i\in\Sbrack{n}}\frac{1}{\delta^2u_i^2}},
  \end{equation}
  then \cref{alg:selection} will select $\widehat{\v}_j$ with high probability.
\end{theorem}

\begin{IEEEproof}
  For convenience, assume that the nodes are indexed such that for $\u\in\CONE^+$, $u_i\leq u_{i+1}$ for $i\in\Sbrack{n-1}$.
Then, the distance from $\u$ to $\BOUNDARY$ using the $\ell_2-\text{norm}$ is equal to $u_1$.
Thus, for some perturbation $\beps\perp\u$ and $\alpha>0$ such that $\uhat=\alpha\u+\beps$ and $\norm{\beps}_2=\sqrt{1-\alpha^2}$, we require $\norm{\beps}_2<\alpha u_1$ to ensure that $\uhat\in\CONE^+$.

By~\cref{prop:alignment}, for some constant $C_1$, one can see that $$\alpha=\cos\theta\Paren{\u,\uhat}>\sqrt{1-\frac{C_1^2}{\delta^2m}}.$$
Then, if ${\alpha^2}/{{(1-\alpha^2)}}>{1}/{u_1^2}$, the desired bound on $\norm{\beps}_2$ is attained.
This condition is equivalent to
\begin{equation*}\label{eq:selection-proof-condition}
  \frac{\delta^2m}{C_1^2} > 1 + \frac{1}{u_1^2}.
\end{equation*}
Noting that the term ${1}/{u_1^2}>n$ dominates, applying~\cref{prop:alignment} yields the sampling requirement~\eqref{eq:finite-selection-requirement}.
\end{IEEEproof}

\vspace{1mm}
Given the discussion in~\cref{sec:prelim:graphs} establishing $\u\in\CONE$, it is intuitive that an eigenvector that is well-centered in $\CONE$ is unlikely to be perturbed beyond $\BOUNDARY$.
Hence, one would anticipate the sample complexity of selecting the correct eigenvector to be large whenever $\u$ contains small entries (in absolute value). 
Theorem~\ref{thm:selection} reveals that this is indeed the case, unveiling the specific functional form of this dependence.
Putting it differently, one expects a \emph{delocalized} eigenvector centrality where $u_i\approx {1}/{\sqrt{n}}$ for all $i\in\Sbrack{n}$ to be ideal for~\cref{prob:selection}, while one with many nodes taking small centrality values -- \ie a localized eigenvector centrality -- to yield a difficult instance of~\cref{prob:selection}.
Additionally, Theorem~\ref{thm:selection} displays the relation between the sample complexity and the relevant eigengap $\delta$. This is as expected, since an eigenvector that is poorly separated from its neighboring eigenvectors in the spectral domain should require more signals for estimation.

\begin{remark}\label{remark:selection}
  \Cref{thm:selection} characterizes the difficulty of~\cref{prob:selection} in terms of the minimum entry of $\u$, providing the sampling requirement to ensure $\uhat\in\CONE^+$.
  Although this condition guarantees that~\cref{alg:selection} will select $\uhat$, it is not a necessary condition.
  That is, if \emph{none} of the eigenvectors of $\Cym$ lie in $\CONE$,~\cref{alg:selection} will pick the one nearest to $\BOUNDARY$.
  Hence, considering the distributions of node centralities within $\u$ and $\Brack{\v_i}_{i=2}^n$ could yield a more precise version of~\eqref{eq:finite-selection-requirement}.
  We draw a practical connection between the difficulty of~\cref{prob:selection} and the \emph{localization} of $\u$ in~\cref{sec:experiments:ws-selection}.
\end{remark}

\section{Numerical Experiments}\label{sec:experiments}

We illustrate the behavior of~\cref{alg:selection} via numerical experiments on two different models of random graphs.
We begin by demonstrating the relationship between sample size and the performance of~\cref{alg:selection} on an \ER graph, then proceed to investigate the influence of the underlying graph structure on the difficulty of~\cref{prob:selection} via simulations on \WS graphs.

For both experiments, a graph filter is excited with $m$ samples of white, Gaussian noise to generate the observed graph signals. 
The probability of selecting the optimal eigenvector centrality estimate over several trials is evaluated. 
For this evaluation, we define the optimal choice as the eigenvector of $\Cym$ that has the greatest inner product with the true eigenvector centrality $\u$.
The graph filters used include a ``square-root'' filter, where $\cH\Paren{\lambda}=\sqrt{\lambda}$, and a ``squared'' filter, where $\cH\Paren{\lambda}=\lambda^2$, both applied to the spectrum of the adjacency matrix, previously scaled and shifted to lie in the interval $\Sbrack{0,1}$.
The square-root filter tends decreases the gap between the dominant eigenvalue and the lower spectrum, leading to a more difficult instance of~\cref{prob:selection}, while the squared filter widens this gap, making selection easier.
Additionally, we consider high-pass versions of these filters, where the spectrum is reversed, \ie $\cH_{HPF}\Paren{\lambda}=1-\cH_{LPF}\Paren{\lambda}$.

\vspace{-1mm}

\subsection{Effect of different graph filters}\label{sec:experiments:erdos-renyi}

To demonstrate the relationship between the number of samples $m$ and the performance of~\cref{alg:selection} for different graph filters, we consider an \ER graph with $n=100$ and $p={(\log n)}/{n}$.
The distribution of centralities in this sparse graph is shown in~\cref{fig:experiments:erdos-renyi}{A}.

We excite the previously described square-root and squared filters, along with their high-pass variants, with white, Gaussian noise.
For each sample size $m$ we evaluate if the optimal estimate of $\u$ was selected.
The minimum centrality in this graph is small, as shown in~\cref{fig:experiments:erdos-renyi}{A}, indicating a difficult instance of~\cref{prob:selection} as expressed by~\cref{thm:selection}.
However, the dominance of the leading eigenvalue of \ER graphs lends itself to a large eigengap $\delta=\lambda_1-\lambda_2$ in this setting.
The influence of this eigengap is demonstrated by the difference in performance between the square-root and squared filters shown in~\cref{fig:experiments:erdos-renyi}{B}: the larger value of $\delta$ from the squared filter results in an easier selection problem.
Additionally, as mentioned in~\cref{remark:selection}, the minimum centrality only dictates the number of samples needed to ensure that the perturbation of $\u$ remains in $\CONE$.
For the purpose of optimal selection, this is overly restrictive, since we only require the perturbation of $\u$ to be closer to $\CONE$ than any other eigenvector of $\Cym$.
Since $\u$ is not highly localized, it is expected that~\cref{alg:selection} will perform well in this scenario for reasonably large values of $\delta$.

\vspace{-1mm}

\subsection{Interplay between eigengap and localization}\label{sec:experiments:ws-selection}

\begin{table}
  \centering
  \caption{Eigengaps of population covariance matrices for Watts-Strogatz graphs}
  \label{tab:experiments:ws-eigengaps}
  \begin{tabular}{ccc}
    \toprule
    $p$ & $\Mean\Sbrack{\delta}$ & $\Var\Sbrack{\delta}$ \\
    \midrule
    \num{0} & \num{1.97e-4} & \num{0} \\
    \num{1e-3} & \num{2.29e-3} & \num{4.54e-6} \\
    \num{1e-2} & \num{4.31e-3} & \num{7.93e-6} \\
    \num{1e-1} & \num{2.03e-2} & \num{2.90e-5} \\
    \num{1} & \num{1.27e-1} & \num{1.12e-4} \\
    \bottomrule
  \end{tabular}
\vspace{-5mm}
\end{table}

We evaluate the performance of~\cref{alg:selection} on graphs drawn from the \WS random graph model.
The parameter $p$, indicating the rewiring probability, has a strong impact on the eigenvector centrality, as illustrated in~\cref{fig:experiments:ws-localization}.
When $p=0$, the \WS is a deterministic, $k$-regular graph, and thus has a constant eigenvector centrality.
The constant vector lies ``in the center'' of $\CONE$, and thus should be the most robust to perturbations due to finite sampling \cf{\cref{remark:selection}}, disregarding the influence of the eigengap $\delta$ in~\eqref{eq:finite-selection-requirement}.
When $0<p\ll 1$, we observe a distinct localization phenomena.
That is, the nodes attached to rewired edges have high centrality, while the centrality of nodes far from a rewired edge have low centrality.
As $p\to 1$, the average distance from rewired connections diminishes, leading to the nodes generally having centrality close to ${1}/{\sqrt{n}}$.
The extreme case of this occurs when $p=1$, where the complete randomness of the graph model yields a highly delocalized eigenvector centrality.

We evaluate the joint influence of the centrality structure and the relevant eigengap on the performance of~\cref{alg:selection}.
We generate our samples through the square-root filter previously described.
Given a collection of $m$ such graph signals, we evaluate the performance of~\cref{alg:selection} in selecting the optimal eigenvector centrality estimate.
The results for \WS graphs drawn $100$ times each from models with $n=500, k=4, \text{ and } p \in \{0,0.001,0.01,0.1,1\}$ are shown in~\cref{fig:experiments:erdos-renyi}{C}.
Notably, except for the scenario where the sample size is small, the models with $p \in \{0.1,0\}$ had the highest rates of correct selection.
This can be explained by two factors:
\begin{enumerate*}
\item A sufficiently large rewiring probability yields a delocalized eigenvector, which is better centered in $\CONE$ than a localized one, and is thus less likely to be perturbed beyond $\BOUNDARY$ due to finite sampling effects, and
\item An eigengap $\delta$ that increases with $p$, reducing the sampling requirement~\eqref{eq:finite-selection-requirement}.
\end{enumerate*}
This second characteristic explains the performance of~\cref{alg:selection} when $p=0$.
Despite the eigenvector centrality being as well-suited to the selection problem as possible, a small eigengap yields an ambiguous selection problem when confronted with finite sampling noise.
Furthermore, the models with $p \in \{0.001,0.01\}$ fare even worse, since they have both small eigengaps \emph{and} localized eigenvector centralities.
We record the eigengaps of $\Cy$ in~\cref{tab:experiments:ws-eigengaps}, where this trend of an increasing eigengap with $p$ can be readily observed.

\vspace{-2mm}

\section{Discussion}\label{sec:discussion}

\vspace{-1mm}

In this work, we considered the blind centrality selection problem, where we seek to estimate the eigenvector centrality of a graph without knowledge of the graph itself.
Rather, we observe a set of graph signals shaped by the network structure via a generic linear graph filter.
Leveraging the shared eigenspaces of the covariance of these signals and the network's adjacency matrix, we propose a simple algorithm for selecting an estimate of the eigenvector centrality from the eigendecomposition of the sample covariance matrix.
We characterize the sampling requirements for correctness of this algorithm in terms of the true eigenvector centrality of the graph.
This is then illustrated through numerical experiments.

This work has many avenues for future research.
The blind inference approach forms a rich paradigm for incorporation of statistical and signal processing techniques for problems in network science.
This could include looking at robust estimation procedures, algorithms that incorporate prior knowledge of the graph structure, and applying blind inference methods to other problems, such as graph matching.

\small
\bibliographystyle{IEEEtran}
\bibliography{ref}

\end{document}